\documentclass[lettersize,journal]{IEEEtran}
\usepackage{amsmath,amsfonts,amsthm,bm} 
\usepackage{multirow}
\usepackage{booktabs}
\usepackage{adjustbox}

\usepackage{multirow} 
\usepackage{multicol}
\usepackage{algorithm}
\usepackage{siunitx}
\usepackage{algpseudocode}
\usepackage{adjustbox}
\usepackage{algorithm}
\usepackage{algpseudocode}
\usepackage{xcolor}
\begin{document}

%

        \title{Personalized Elastic Embedding Learning for On-Device Recommendation}
      

%
%
\author{Ruiqi~Zheng,
        Liang~Qu,
        Tong~Chen,~\IEEEmembership{Member,~IEEE},
        Kai~Zheng,~\IEEEmembership{Senior Member,~IEEE},
        Yuhui~Shi*,~\IEEEmembership{Fellow,~IEEE}, and 
        Hongzhi~Yin*,~\IEEEmembership{Senior Member,~IEEE}\IEEEcompsocitemizethanks{\IEEEcompsocthanksitem Ruiqi~Zheng, Liang~Qu, Tong~Chen, Hongzhi~Yin are with School of Information Technology and Electrical Engineering, The University of Queensland, Brisbane, Australia, Email: ruiqi.zheng@uq.net.au, Email: l.qu1@uq.net.au, Email: tong.chen@uq.edu.au, Email: h.yin1@uq.edu.au \IEEEcompsocthanksitem Kai~Zheng is with School of Computer Science and Engineering, University of Electronic Science and Technology of China, China, Email: zhengkai@uestc.edu.cn\IEEEcompsocthanksitem Ruiqi~Zheng, Liang~Qu, Yuhui~Shi are with Department of Computer Science and Engineering, Southern University of Science of Technology, Shenzhen, China, Email: shiyh@sustech.edu.cn}
        \thanks{*Corresponding authors: Hongzhi~Yin and Yuhui~Shi}}
\IEEEtitleabstractindextext{
\begin{abstract}

To address privacy concerns and reduce network latency, there has been a recent trend of compressing cumbersome recommendation models trained on the cloud and deploying compact recommender models to resource-limited devices for the real-time recommendation. Existing solutions generally overlook device heterogeneity and user heterogeneity. They require devices with the same budget to share the same model and assume the available device resources (e.g., memory) are constant, which is not reflective of reality.  Considering device and user heterogeneities as well as dynamic resource constraints, this paper proposes a Personalized Elastic Embedding Learning framework (PEEL) for the on-device recommendation, which generates Personalized Elastic Embeddings (PEEs) for devices with various memory budgets in a once-for-all manner, adapting to new or dynamic budgets, and addressing user preference diversity by assigning personalized embeddings for different groups of users.
Specifically, it pretrains a global embedding table with collected user-item interaction instances and clusters users into groups. Then, it refines the embedding tables with local interaction instances within each group. PEEs are generated from the group-wise embedding blocks and their weights that indicate the contribution of each embedding block to the local recommendation performance. Given a memory budget, PEEL efficiently generates PEEs by selecting embedding blocks with the largest weights, making it adaptable to dynamic memory budgets on devices. Furthermore, a diversity-driven regularizer is implemented to encourage the expressiveness of embedding blocks, and a controller is utilized to optimize the weights. Extensive experiments are conducted on two public datasets, and the results show that PEEL yields superior performance on devices with heterogeneous and dynamic memory budgets.

\end{abstract}


\markboth{Journal of \LaTeX\ Class Files,~Vol.~14, No.~8, August~2021}%
{Shell \MakeLowercase{\textit{et al.}}: A Sample Article Using IEEEtran.cls for IEEE Journals}

    }

\maketitle 
      
\IEEEdisplaynontitleabstractindextext

%
\IEEEpeerreviewmaketitle

\section{Introduction}
%
%

%
%

 


\IEEEPARstart{R}{ecommender} systems (RecSys) \cite{covington2016deep,guo2017deepfm,10.1145/3437963.3441762}  have been widely shown as an effective technique for helping users filter out irrelevant information. In a nutshell, a typical pipeline of contemporary RecSys \cite{10.1145/3437963.3441738,10144391} trains a powerful recommender model with user data on a cloud server, and the cloud-based recommender then handles users' recommendation requests by sending the recommendation results to their devices. 
In this pipeline, latent factor-based models \cite{guo2017deepfm,10.1145/3437963.3441738} are the most prevalent RecSys, which map users/items into embedding representations for pairwise affinity calculation through deep neural networks or similarity metrics \cite{johnson2019billion}. Because of the sheer volume of users and items (e.g., the million-scale video set in social media platform \cite{covington2016deep}), the embedding table is the dominating factor of memory consumption in a RecSys model compared to other parameters like weights and biases \cite{10.1145/3366423.3380170,qu2022single}. Consequently, this conventional RecSys pipeline is admittedly resource-intensive in terms of the need for large-scale storage and communication overhead \cite{10.1145/3511808.3557065}.

With the increasing expectation on timeliness \cite{stoica2017berkeley,10.1145/3534678.3539080} and efficiency \cite{tinyTL,xia2023towards}, there has been a recent trend in moving trained models from central servers to edge devices  (e.g., smartphone and smart car) to reduce reliance on the cloud \cite{10.1145/3580364, xia2022device, 10.1145/3579355}. Unlike cloud servers with abundant memory space and computation resources, edge devices fall short in both memory and computation capacity \cite{10.1145/3450494}. Therefore, memory-efficient RecSys \cite{10.1145/3366423.3380170,10.1145/3366423.3380151,10.1145/3394486.3403059} has emerged to deploy compact models on devices with limited memory budgets. Targeting on compressing the most memory-consuming component of RecSys, i.e., the embedding table, early discrete methods \cite{10.1145/2911451.2911502,10.1145/2339530.2339611} transform embeddings into condensed binary codes. To further preserve model expressiveness with real-valued embeddings,  multi-dimensional embeddings, including heuristics \cite{ginart2021mixed}, automated hyper-parameter selection \cite{joglekar2020neural,liu2020automated}, and embedding pruning \cite{liu2021learnable,10.1145/3477495.3532060}, have been recently studied to assign each user/item a different embedding dimensionality, so as to reduce excessive memory consumption. In a similar vein, we have seen compositional embedding methods with a small set of meta-embeddings \cite{10.1145/3366423.3380151, 10.1145/3394486.3403059}, whose subsets are used to compose unique user/item embeddings.

{
However, most memory-efficient solutions result in the same compact model for all users\footnote{We assume one device for one user in this paper, and denote the devices by their corresponding users.}, neglecting \textbf{device heterogeneity} and \textbf{user heterogeneity}. }
For devices, the ever expanding diversity of IoT facilities has brought considerable variations in device capacities (e.g., memory space and latency). Using a uniformly sized model across heterogeneous devices means that the model has to be configured for the most resource-constrained device, thus being infeasible for large-scale on-device applications \cite{10.1145/3579355}. 

\begin{figure}[t]
\centering
\includegraphics[width=0.47\textwidth]{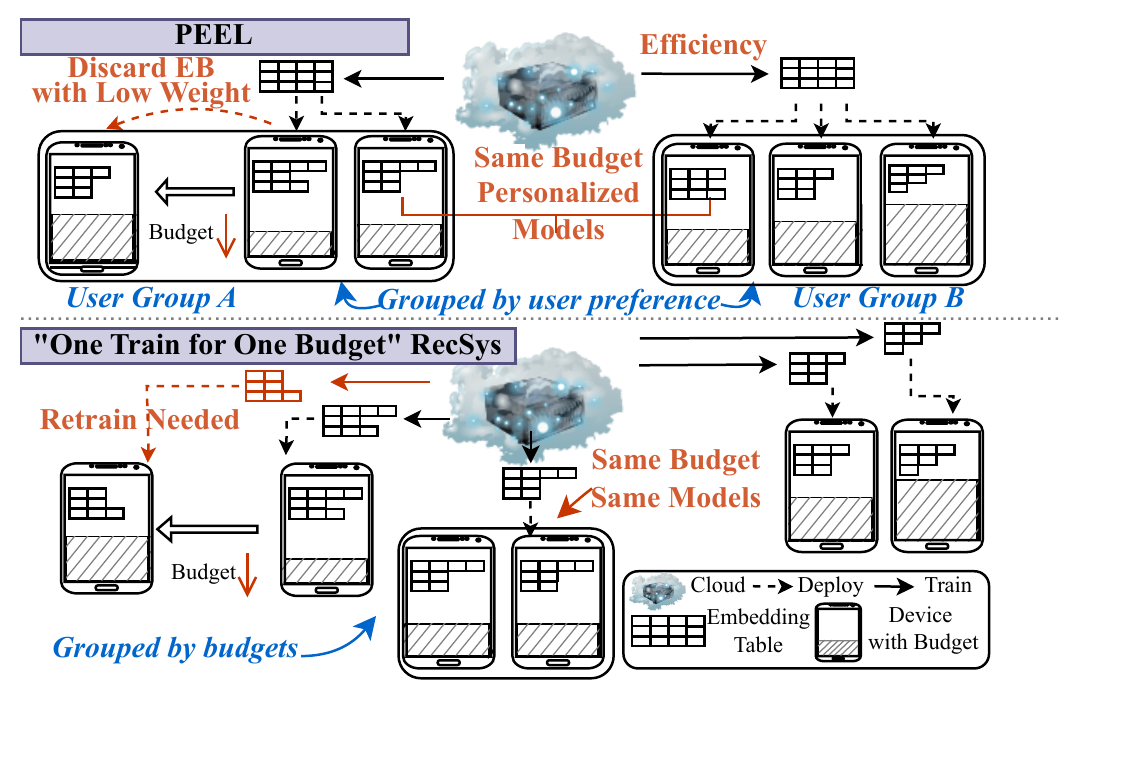} 
\caption{Difference between PEEL and existing ``one train for one budget'' memory-efficient RecSys.}
\label{fig:intro}
\vspace{-15pt}
\end{figure}

{
Although some existing on-device RecSys methods \cite{rockyChenKDD21,han2021deeprec,qu2023continuous, qu2023budgeted} allow a recommender to be customized for each memory budget, allocating different models for distinctive devices, they have largely overlooked the fact that the available resources on one device are always dynamic in real-world scenarios. }Taking smartphones as an example, when background applications consume excessive memory or the battery is running low, the complexity of the deployed on-device recommender has to be further reduced to keep operating. Even with automated embedding customization methods \cite{joglekar2020neural,liu2021learnable,10.1145/3477495.3532060} that avoid rebuilding a model from scratch (e.g. \cite{rockyChenKDD21,Yu_Gong_He_Zhu_Liu_Ou_An_2021}), it is still impractical to lodge a new recommender customization request each time a new memory allocation is required.

{
For user heterogeneity, as users typically exhibit a variety of behavioral patterns, the distribution of a particular user may deviate from the global distribution \cite{10.1145/3534678.3539263, 10.1145/3437963.3441835}.}
On this occasion, 
adopting a globally shared model configuration will inevitably impair the recommendation quality for individual users. To account for heterogeneity among users, a straightforward solution is to develop a personalized model with local instances on every device \cite{10.1145/3534678.3539263}. This, however, is far from being cost-effective given the scale of users in modern e-commerce sites \cite{10.1145/3477495.3532060}. Moreover, training isolated on-device recommenders are subject to strong data scarcity, and introduces additional resource consumption on top of the inference process. 
A remedy to this is to enable cloud-device collaborations \cite{10.1145/3447548.3467097, 10.1145/3555374} to achieve personalized on-device models given a centralized cloud model. Unfortunately, such methods fail to adapt to varying device-specific memory budgets, and the back-and-forth communications required for training are time-consuming and inflexible when facing new users/devices. 



In light of the challenges, we propose a sustainable pipeline for deploying lightweight recommenders on-device without the need for reconfiguring or retraining the whole model for every possible use case, {addressing both device and user heterogeneity}. Specifically, on the central server, we optimize user and item embeddings by jointly learning all user-item interactions and the importance of different embedding segments, and a lightweight item embedding table\footnote{In practice, one device carries all items' embeddings while only needs to store its corresponding user's embedding, so we mainly focus on item embedding compression.} can be efficiently tailored by selectively picking the most weighted segments w.r.t. any given memory budget thereafter. We term those compressible embeddings Personalized Elastic Embeddings (PEEs). 
Rather than creating a single, high-dimensional embedding, we segment each full embedding into smaller chunks, namely embedding blocks, of which the importance weights are learned via an end-to-end bi-level optimization scheme. PEE is distinct from compositional embeddings \cite{10.1145/3366423.3380151,10.1145/3394486.3403059} as 
embedding blocks for each item are not shared and exclusive, promoting stronger
expressiveness. As shown in Figure \ref{fig:intro}, neither retraining nor reconfiguration is needed to develop personalized models for users with heterogeneous memory budgets. Besides, once the memory budget decreases after deployment, an on-device model can instantly deactivate less important blocks, thus retaining its accuracy under the constantly changing device capacity.

Furthermore, to assign each user a personalized embedding table that can best represent her personal interest while avoiding the hefty local training and cloud-device communication overhead, we propose a group-level fine-tuning strategy. In short, we cluster users into groups based on their pretrained embeddings, and further refine the globally learned item embeddings with instances from the same user group. In this way, a full item embedding table can be optimized for each specific user group when the block-wise importance weights are being learned. 
On the one hand, compared with ``one model for all'' methods, the group-level fine-tuning strategy serves as a denoising step that enables the information learned in each group-specific embedding table to be more focused. On the other hand, compared with ``one model for each'' methods, our strategy allows for data augmentation by herding similar users, where the trade-off between specificity and generalizability can be easily controlled by the granularity of user groups. In addition, our group-level fine-tuning is executed on the server side in a one-off manner, and can handle all user groups in parallel.


To this end, we propose PEEL, namely the \underline{P}ersonalized \underline{E}lastic \underline{E}mbedding \underline{L}earning (\textbf{PEEL}) framework that identifies the optimal PEEs for on-device recommendation under varying memory budgets in one shot. 
In PEEL, we train a controller to quantify the contribution from each embedding block based on the recommendation loss and item popularity w.r.t. every user group, promoting personalization. The fine-tuning process and controller optimization are performed alternately via gradient descent, which results in a unique item embedding table and a set of block-wise importance scores for each user group to facilitate the subsequent PEE deployment. Meanwhile, a parameter-free similarity function is utilized for efficient on-device inference with the learned elastic embeddings to generate final recommendations. 
The contributions of this paper are threefold: 
\begin{itemize}
\item We propose a group-level fine-tuning strategy to develop personalized models for memory-efficient recommender systems, when facing the user and device heterogeneity. By synthesizing PEEs composed of embedding blocks and importance weights, our strategy generates RecSys for heterogeneous and dynamic memory budgets without retraining or reconfiguring the on-device model.
\item We propose PEEL, a new framework to address the essential obstacles associated with implementing this novel strategy. PEEL can produce quality embedding blocks with importance weights learned via a bi-level optimization scheme, to support PEE deployment under heterogeneous and dynamic memory budgets effectively and efficiently.
\item Extensive experiments are conducted on two real-world datasets under various memory budgets. Experimental results demonstrate that PEEL outperforms contemporary leading methods. 
\end{itemize}

The rest of the paper is organized as follows. Section \ref{sec:related work} introduces the related work. Section \ref{sec:pre} provides preliminaries and problem formulation. Section \ref{sec:proposed method} will explain the details of the proposed framework. The experiments are introduced in Section \ref{sec:experiment}, followed by a conclusion in Section \ref{sec:conclusion}.


\section{Related Work}
In this section, we review recent literature on related areas including model compression and on-device recommendation.
\label{sec:related work}
\subsection{Model Compression}
With the increasing concerns for timeliness and privacy, there has been a recent trend in moving models from cloud servers to edge devices. The fundamental challenge is fitting the complex convolutional deep neural network (DNN) models to resource-constrained devices. Model compression techniques are proposed to reduce the model sizes while making the slightest compromise with the model accuracy, which can be generally divided into four categories. (1) \textit{Low-rank Factorization:} It factorizes a large matrix into smaller matrices. The final
weight layer \cite{sainath2013low}, and convolutional layers \cite{denton2014exploiting,lebedev2014speeding} can be factorized for model compression. (2) \textit{Knowledge Distillation:} The complex teacher model trained on a large dataset is transferred to a lightweight student model to generalize to unseen data, and reduce the model size \cite{hinton2015distilling,romero2014fitnets}. (3) \textit{Quantization:} The DNN models can be shrunk by using fewer bits to represent the activations and weights, including dense layer quantization \cite{gong2014compressing}, weight binarization \cite{courbariaux2015binaryconnect}, and bit-wise operation \cite{kim2018efficient}. (4) \textit{Pruning:} Weight pruning \cite{han2015learning}, neuron pruning \cite{srinivas2015data}, filter pruning \cite{li2017pruning}, and layer pruning \cite{chen2018shallowing} remove the lesser important connections for storage reduction. Despite the effectiveness of these DNN model compression methods, they are not necessarily feasible to generate a memory-efficient RecSys, since the embedding table is the primary factor of memory consumption compared to parameters such as biases and weights.
\subsection{On-Device Recommendation}
Recently, on-device recommender systems emerged in response to users’ needs for low latency recommendations and have demonstrated desired performance with tiny model sizes. On-device recommender systems should be memory-efficient to fit the device memory budgets. Specifically, hashing techniques are implemented to transform embedding tables into condensed binary codes \cite{10.1145/2911451.2911502,10.1145/2339530.2339611}. Compositional embedding methods \cite{10.1145/3366423.3380151,10.1145/3394486.3403059,liang2023learning} develop a set of meta-embeddings, whose subsets compose user/item embeddings to shrink the memory consumption of embeddings. Multi-dimensional embeddings, including heuristic \cite{ginart2021mixed}, hyper-parameter optimization \cite{joglekar2020neural,liu2020automated}, and embedding pruning \cite{liu2021learnable,10.1145/3477495.3532060}, assign embedding dimensions to each feature automatically to reduce excessive memory consumption. RULE \cite{rockyChenKDD21} utilizes elastic embedding and evolutionary search to customize a lightweight RecSys under budgets. Session-based RecSys \cite{changmai2019device,xia2023towards} is proposed to use users' real-time intentions to generate on-device recommendations. DeepRec \cite{han2021deeprec} uses embedding sparsification and model pruning techniques to conduct the model training under a small memory budget and the model fine-tuning with local user instances. LLRec \cite{wang2020next} utilizes tensor-train factorization to compress the embedding in the POI recommendation scenarios. TT-Rec \cite{yin2021tt} uses tensor-train decomposition for model compression at the embedding layer level. 
Although existing on-device RecSys methods show promising results, they neglect device and user device heterogeneity. In contrast, our design of PEEL deploys lightweight recommenders on-device without the need for reconfiguring or retraining.

\section{{Preliminaries and Problem Formulation}}
\label{sec:pre}
{We provide definitions for key concepts, namely embedding block and personalized elastic embedding for our proposed method. Additionally, we present the problem formulation. The frequently used notations are shown in Table \ref{tab:sign}. }

    \textbf{Definition 1} Embedding Block (EB): The full embedding $\textbf{v}_j \in \mathbb{R}^{D}$ for an item 
${v}_j \in \mathcal{V}$ consists of $N$ separate embedding blocks $ \textbf{e}_j^n \in \mathbb{R}^d (1\leq n \leq N)$, where $D=Nd$ and $\textbf{v}_j = [\textbf{e}_j^1,\textbf{e}_j^2,\cdot\cdot\cdot,\textbf{e}^N_j]$. An embedding block $\textbf{e}_j^n$ is a $d-$dimensional vector.

    \textbf{Definition 2} Personalized Elastic Embedding (PEE): For a user $u_i \in \mathcal{U}$, the personalized elastic embedding of item $v_j$ is represented as $\textbf{v}^{i}_{j}=[\textbf{e}^n_j]_{n \in \mathcal{S}_{v_j}^{u_i}}$, the concatenation of the selected embedding blocks, where set $\mathcal{S}_{j}^{i}$ stores the indexes of selected embedding blocks, and $|\mathcal{S}_{j}^{i}| \leq N$. For each user, $\mathcal{S}_{j}^{i}$ will vary across $v_j \in \mathcal{V}$, stating that all items' elastic embeddings are personalized.

\textbf{Problem 1:} On-Device Recommendation with Personalized Elastic
Embeddings: For user $u_i \in \mathcal{U}$ under on-device memory budget $M$, PEEL searches $\{\mathcal{S}^{i}_{j}\}_{v_j \in |\mathcal{V}|}$ to obtain personalized elastic embeddings for all items for accurate recommendations, satisfying the restriction $size( \{\textbf{v}_{j}^i\}_{1 \leq j \leq |\mathcal{V}|} \cup \{{\textbf{u}_i} \}) \leq M$. Since one device carries all items’ embeddings while only needing to store its user’s embedding, we mainly focus on item embedding compression.

\begin{table}[htbp]
  \centering
  \caption{Frequently used notations.}
    \begin{tabular}{ll}
    \toprule
    \textbf{Notations} & \textbf{Descriptions} \\
    \midrule
    \midrule
    $\mathcal{U}/\mathcal{V}$ & the set of users/items \\
    $\textbf{v}_j $ & the full embedding for item $v_j$ \\
    $\textbf{v}_j^i $ & the PEE for item $v_j$ of user $u_i$\\
    
    $\textbf{e}^n_j$ & the $n^{th}$ EB for the full embedding $\textbf{v}_j$ \\
    $\textbf{u}_i$  & the embedding for user $u_i$ \\
    $\textbf{E}^{n}_{G_v}$  & the $n^{th} $embedding block for ${G_v}^{th}$ item group\\
    $\textbf{E}^{n}$  & the collection of $n^{th} $ EB for all item groups\\
    $[\textbf{e}_j^{n}]_{n\in {n_1,n_2}}$ & the vector concatenation of $\textbf{e}_j^{n_1}$ and $\textbf{e}_j^{n_2}$ \\
    $\{\textbf{e}_j^{n}\}_{n\in {n_1,n_2}}$ & the set of $\textbf{e}_j^{n_1}$ and $\textbf{e}_j^{n_2}$ \\
    $\mathcal{S}^{i}_{j}$ & the set of EB indexes for item $v_j$ w.r.t. user $u_i$\\
    $\mathcal{G}_{u_i}$ & the set including the neighbors of user ${u_i}$ \\
  $\mathcal{D}$ &
     the set of all training samples \\
    $|\mathcal{S}|$ & the size of set $\mathcal{S}$ \\
    $N$ & the number of EBs for one item \\
    $C$ & the maximum number of selected EBs \\
    $d$ & the length of embedding block $\textbf{e}^n_j$ \\
    $l$ & the index of neural network layer\\
    $D$ & the length of full embedding $\textbf{v}_j$ \\
    $M$ & the on-device memory budget \\
    $size(\textbf{v}_i)$ & the memory consumption of embedding $\textbf{v}_i$ \\ 
    \bottomrule
    \end{tabular}%
  \label{tab:sign}%
\end{table}%

\section{Proposed Method}
\label{sec:proposed method}
\begin{figure*}[t]
\centering
\includegraphics[width=1\textwidth]{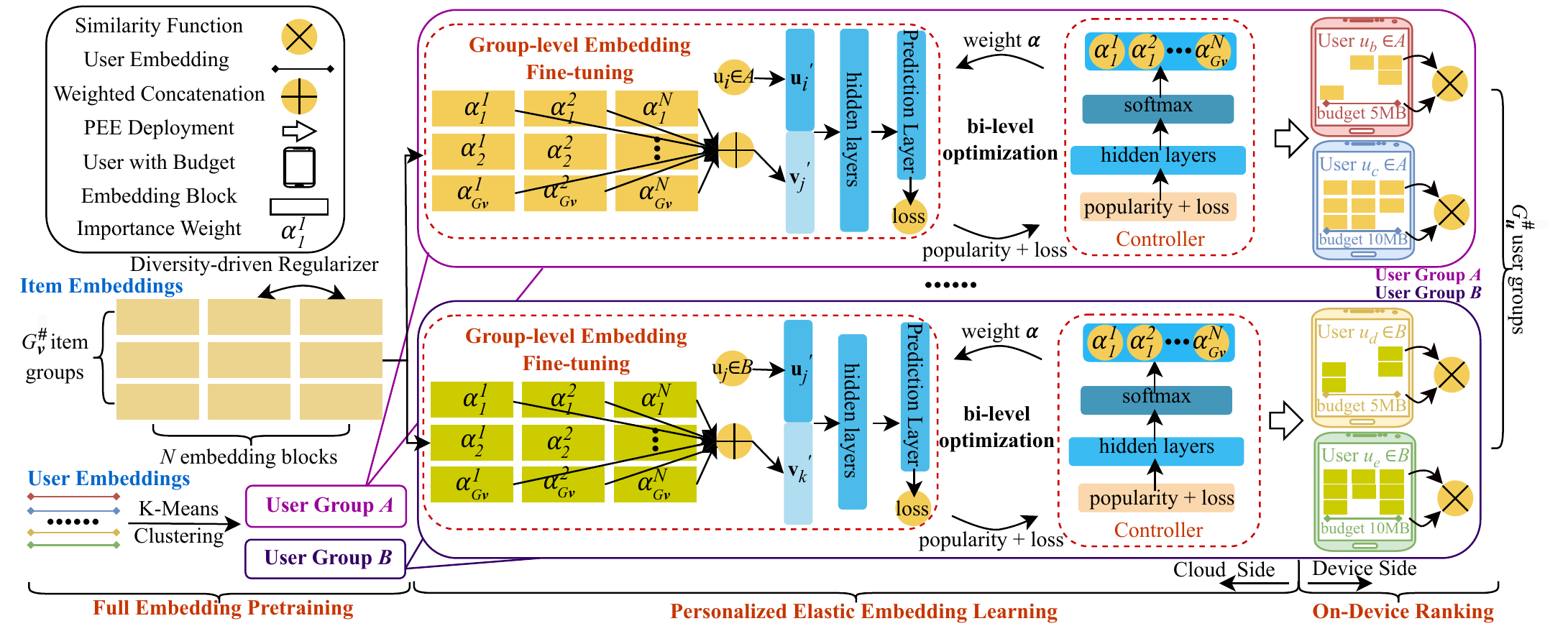} 
\caption{The overview of PEEL. a) The pretraining stage trains models with collected data instances
a diversity-driven regularizer. b) the PEE learning stage refines models with local data instances within the user group, and a controller learns weights for embedding blocks. c) The on-device ranking stage outputs the recommendations with a parameter-free similarity function.}
\label{fig:overview}
\vspace{-10pt}
\end{figure*}

{In this section, we introduce the search space and our proposed method PEEL}. As shown in Figure \ref{fig:overview}, PEEL has three main components: full embedding pretraining, personalized elastic embedding learning, and on-device ranking. Concretely, in the pretraining stage, we pretrain recommendation models with collected data instances and utilize a diversity-driven regularizer to output the item embedding table. Users are clustered into groups based on user embeddings. In the PEE learning stage, for every user group, a copy of the item embedding table is refined based on the local data instances within the user group, and a controller learns weights for embedding blocks, indicating the contributions of each embedding block to the local recommendation performance. The interdependent embedding table refining and weights optimization are conducted in a bi-level optimization manner, instead of a conventional attention mechanism to prevent overfitting. Finally, in the on-device ranking stage. The learned PEEs for all items and only one specific user embedding are deployed on the corresponding device. A parameter-free similarity function is implemented to rank all the items and output the recommendations.

\subsection{Search Space of PEEL}
Based on the definition in Section \ref{sec:pre}, different embedding blocks are designated for all items and users. Assuming each personalized elastic embedding incorporates at least one block, the complexity of the search space for all items would be $O((2^{N}-1)^{|\mathcal{V}|})$, and expends to $O({(2^{N}-1)^{|\mathcal{V}||\mathcal{U}|}})$ for all users, where $|\mathcal{V}|$ and $|\mathcal{U}|$ represent the total number of items and users, respectively, which can reach into the millions in real-world scenarios \cite{10.1145/3477495.3532060}. To manage this complexity and enhance local instance relevance, a group-level fine-tuning strategy is employed, clustering users into $G^{\#}_{u}$ groups as elaborated in Section \ref{sec:method_user_group}. It is worth noticing that the number of users in each group
is not necessarily the same, since the clustering procedure
is not aimed at splitting users evenly, but at identifying users with
similar properties for personalization.

For item segmentation, the aim is to condense the search space efficiently. This is achieved by evenly dividing items into $G^{\#}_v$ groups. Within this framework, users belonging to the same group $\mathcal{U}_{G_u} (1 \leq G_u \leq G^{\#}_{u})$ share a common embedding block selection approach $\mathcal{S}^{u}_{v}$ for items in the same group $\mathcal{V}_{G_v} (1 \leq G_v \leq G^{\#}_{v})$. Popularity-based segmentation is adopted for items, recognizing that item popularity significantly influences recommendation quality. Popular items, which typically have extensive interaction records, benefit from a higher dimensionality in embedding (i.e., more blocks), enabling a more nuanced representation of associated contexts. Conversely, items with fewer interactions require fewer blocks. To implement this, all items $v_i \in \mathcal{V}$ are sorted in descending order by popularity. The items are then divided into $G^{\#}_{v}$ groups through equal segmentation of this ordered set. The complexity of search space is hence substantially reduced to $O({(2^{N}-1)^{G^{\#}_{v}G^{\#}_{u}}})$ with number of item groups $G^{\#}_{v} \ll |\mathcal{V}|$ and number of user groups $G^{\#}_{u} \ll |\mathcal{U}|$.

\subsection{Embedding Pretraining}
Instead of directly optimizing the PEEs for different users under an on-device budget, we learn the full embedding of all items collaboratively by classical recommender systems on the cloud side to comprehensively utilize the user-item interactions and cluster users with different statistical distributions. Furthermore, to ensure the diversity and expressiveness of each embedding block, a diversity-driven regularizer is applied in the full embedding pretraining process.

\subsubsection{Base Recommender Systems}
It is worth noticing that our contributions focus on the PEE learning framework. This process can be accomplished by any latent model, from deep recommender systems \cite{10.1145/3038912.3052569} to matrix factorization \cite{5197422}.
we employ a graph neural network model, LightGCN \cite{he2020lightgcn}, as the base recommender system since it shows competitive performance under different scenarios \cite{zhu2022bars}. LightGCN treats each user/item as a node and the interactions as edges in a bipartite graph. The embeddings of users and items are updated by propagating their neighbors :

\begin{equation}
\begin{aligned}
\eta_{ij} &= (\sqrt{|\mathcal{G}_{u_i}| \cdot |\mathcal{G}_{v_j}|} \ \  )^{-1} \\
\textbf{v}^{l}_{j} &= \sum\limits_{i \in \mathcal{G}_{v_j}}  \eta_{ij} \textbf{u}_i^{l-1} \\
\textbf{u}^{l}_{i} &= \sum\limits_{j \in \mathcal{G}_{u_i}}  \eta_{ij} \textbf{v}_j^{l-1}
\end{aligned}
\end{equation}
where $\mathcal{G}_{u_i}$ and $\mathcal{G}_{v_j}$ represent the neighbor set for user $u_i$ and item $v_j$, respectively. $\eta_{ij}$ is the graph laplacian normalization term \cite{rao2015collaborative} and $1 \leq l \leq L_1$. We take the final layer $L_1$ of LightGCN as the full embeddings for users and items by $\textbf{u}_i = \textbf{u}_i^{L_1}$ and $\textbf{v}_j = \textbf{v}_j^{L_1}$, respectively. It is worth noting that the choice of only using the final layer of the LightGCN is due to the design of PEEL, as it incorporates multiple embedding blocks, and a diversity-driven regularize. The empirical results are shown in Section \ref{sec:ablation}.


\subsubsection{Diversity-driven Regularizer}
Unlike traditional recommender systems, user/item embeddings in PEEL consist of several embedding blocks. To maximize the information diversity of each PEE, the embedding blocks should be as distinctive as possible from each other. In this manner, PEE learning can be more effective since it avoids selecting similar embedding blocks. The loss function $\mathcal{L}_{rec}$ of PEEL has been modified as follows by adding a diversity-driven regularizer after the original recommendation loss:
\begin{equation}
\begin{aligned}
\mathcal{L}_{rec} =& - \sum_{u_i,v_j,v_k \in \mathcal{D}} ln\sigma(\hat{y}_{ij} - \hat{y}_{ik}) \\ & - 
\lambda \sum^{N}_{n_1=1}\sum^{N}_{n_2=n_1+1} || \textbf{E}^{n_1} - \textbf{E}^{n_2} ||^{2}_{F}
\end{aligned}
\label{equa:loss_funcion}
\end{equation}
where the first term is Bayesian Personalized Ranking (BPR) loss \cite{BPR}, and the second term is the diversity-driven regularizer. Set $\mathcal{D} = \{(u_i,v_j,v_k) | \ v_j \in \mathcal{V}^{+}_{u_i} \wedge v_k \in \mathcal{V}\setminus\mathcal{V}^{+}_{u_i}\}$ consists of all training samples, where $\mathcal{V}^{+}_{u_i}$ denotes the set of items user $u_i$ likes, and $\mathcal{V}\setminus\mathcal{V}^{+}_{u_i}$ excludes $\mathcal{V}^{+}_{u_i}$ from $\mathcal{V}$. $\hat{y}_{ij}$ and $\hat{y}_{ik}$ represent the prediction results of user $u_i$ to item $v_j$ and item $v_k$, repectively. As for the diversity-driven regularizer, 
$\textbf{E}^{n}$ denotes the sets of the $n^{th}$ embedding blocks for all item groups
and $|| \cdot ||^{2}_{F}$ represents the squared Frobenius norm \cite{Frobenius_norm}.  In this manner, two embedding blocks for the same item group are encouraged to be distinctive from the element-wise level, promoting the diversity of embedding blocks.

\subsection{Personalized Elastic Embedding Learning}
\label{sec:method_user_group}

\subsubsection{Group-Level Fine-Tuning Strategy}
As shown in Figure \ref{fig:overview}, the full embedding learning procedure outputs embedding table $\textbf{E} $, which consists of $G_{\textbf{v}}^{\#} \times N $ embedding blocks. However, users have distinctive instance distributions and even hold diversified preferences toward the same item. Therefore, Given a set of user embeddings, we utilize $K$-Means \cite{ahmed2020k} to divide users into set $\mathcal{U'} = \{\mathcal{U}_1, \mathcal{U}_2, \cdot\cdot\cdot, \mathcal{U}_{G_{u}^{\#}} \}$ to minimize the within-cluster variance:
\begin{equation}
     \arg \min_{\mathcal{U'}} \sum^{G_{u}^{\#}}_{i=1} \sum_{\textbf{x} \in \mathcal{U}_{i}} || \textbf{x} - \boldsymbol{\mu}_{i}||^{2}
\end{equation}
where $\boldsymbol{\mu}_{i}$ is the mean of embeddings in $\mathcal{U}_i$. $E_{G_u}$ represents the input embedding table for ${G_u}^{th}$ user group. Initially $E_{G_u} = \textbf{E}$, for $1 \leq G_u \leq G_{u}^{\#}$. User-item data instances $(u_i, v_j)$ will be used for learning within the user group $\mathcal{U}_{G_u}$, only when the user belongs to that group (i.e., $u_i \in \mathcal{U}_{G_u}$). For all user groups, the personalized elastic embedding learning procedure is executed on the server side in a one-off
manner, and can handle all user groups in parallel. We illustrate one PEE learning procedure within a user group, and omit the user group index $G_u$ for notation simplification.

Within the $n^{th}$ embedding block for ${G_v}^{th}$ item groups $\textbf{E}^{n}_{G_v} \in \mathbb{R}^{\frac{|\mathcal{V}|}{G^{\#}_{\textbf{v}}} \times d}$, $e^{n}_j \in \mathbb{R}^{d}$ is a $d$-dimensional embedding block for item $v_j$. Once the user-item data instance $(u_i, v_j)$ is fed into the RecSys, the concatenation of embedding blocks $[\textbf{e}^{n}_j]_{1 \leq n \leq N}$ generates the item embedding $\textbf{v}'_j$. However, in practice, the magnitude of the $d-$dimensional vectors $\textbf{e}^{n}_j$ from different embedding blocks may vary significantly, which leads to one vector dominating others. To address this issue, Batch-Norm \cite{ioffe2015batch} and Tanh activation \cite{karlik2011performance} are utilized to transform the embeddings:

\begin{equation}
    \hat{\textbf{e}}_j^{n} = tanh(\frac{\textbf{e}^{n}_{j}-\mu_\mathcal{B}^{n}}{\sqrt{(\sigma^{n}_{\mathcal{B}})^{2} + \epsilon}})
    \label{equa:normalization}
\end{equation}
where $1 \leq n \leq N$, $(\sigma^{n}_{\mathcal{B}})^{2}$ represents mini-batch variance, $\mu_\mathcal{B}^{n}$ represents mini-batch mean, and $\epsilon$ is a small constant for arithmetic stability. The activation function maps the normalized embeddings between 0 and 1. After transformation, we get $\textbf{v}^{'}_j = [\hat{\textbf{e}}^{n}_j]_{1 \leq n \leq N}$ with magnitude-comparable embedding.

Aiming to select proper embedding blocks, the importance weights $\boldsymbol{\alpha} = \{\alpha^i_j\}_{1 \leq i \leq N, 1 \leq j \leq G_{v}^{\#}}$  are utilized to reflect the contribution of each embedding block to the local recommendation performance. Then the embedding blocks with large importance weights are retained during the PEE deployment under particular on-device budgets. For the $n^{th}$ embedding block for ${G_v}^{th}$ item group $\textbf{E}^{n}_{G_v}$, the associated importance weight is $\alpha^{n}_{G_v}$, and every $d-$dimensional vector within one embedding block shares the same importance weight. The importance weights are generated by a controller, which will be introduced in Section \ref{sec:controller}. The user embeddings are initialized by $\{\textbf{u}_i \}_{1 \leq i \leq |\mathcal{U}|}$, the output of pretraining process. The item embedding $\textbf{\textbf{v}}'_j = [\alpha^{n}_{G_{v}} \cdot \hat{\textbf{e}}^{n}_j]_{1 \leq n \leq N} \in \mathbb{R}^{D}$ when item $v_j$ belongs to the ${G_{v}}^{th}$ item group.

User and item representations are concatenated together as $\textbf{h}_0 = [\textbf{u}_i, \textbf{v}'_j]$, the input of $L_2$ fully-connected hidden layers:
\begin{equation}
    \textbf{h}_{l} = tanh(\textbf{W}_{l-1}\textbf{h}_{l-1}+\textbf{b}_{l-1})
\end{equation}
where $\textbf{W}_{l-1}$ is the weight, $tanh(\cdot)$ is the activation function, $\textbf{h}_{l-1}$ is the outputs of the previous layer, and $\textbf{b}_{l-1}$ is the bias. Finally, The prediction layer calculates the prediction $\hat{y}$:
\begin{equation}
\hat{y} = \sigma(\textbf{W}_o \textbf{h}_{L_2} + \textbf{b}_o)
\end{equation}
where $\textbf{b}_o$ is the bias, and $\textbf{W}_o$ is the weight. The specific recommendation task determines the choices of the activation function $\sigma(\cdot)$. In this manner, the refined RecSys with embedding blocks updates the neural network parameters and refines the embedding tables, by training on the user-item instances within the corresponding user group.

\subsubsection{Controller}
\label{sec:controller}
As shown in Figure \ref{fig:overview}, to evaluate the importance of each embedding block, importance weights $\boldsymbol{\alpha}$ are associated with corresponding embedding blocks, revealing their contribution to the final recommendation performance. Since it is impractical to manually set the importance weights for all EBs, we put forward a parallel learning task for $\boldsymbol{\alpha}$ in PEEL when fine-tuning all embeddings. However, straightforwardly learning the importance weights via conventional attention mechanisms is a less effective solution. 
This is because the optimization of those weights and embedding tables are highly interdependent, which tends to overfit the training data and unable to generalize \cite{10.1145/3397271.3401436}. Therefore, we propose a controller to learn the weights. Given a particular user-item interaction instance, the fine-tuned RecSys with EBs propagates the instance to obtain the loss value. Then, the context information (i.e., the item's popularity and gained loss value) is fed into $L_3$ fully-connected hidden layers to acquire hierarchical feature representations. The input $\textbf{h}_0^{c} = [p_1, \cdot\cdot\cdot, p_{G^{\#}_{v}}, \mathcal{L}_{rec}]$, and $p_{G_v}$ represents the portion of one item group above all local item groups.  After that, output vector $\textbf{e}_c \in \mathbb{R}^{|\boldsymbol{\alpha}|}$ is generated by the output layer with softmax activation function:

\begin{equation}
\begin{aligned}
    \textbf{h}^{c}_{l} &= tanh(\textbf{W}_{l-1}^{c}\textbf{h}^{c}_{l-1}+\textbf{b}^{c}_{l-1})
    \\
    \textbf{e}_c &= softmax(\textbf{W}^{c}_o \textbf{h}_{L_2} + \textbf{b}^{c}_o)
    \end{aligned}
\end{equation}
Each importance weight is assigned by its corresponding value in the output vector, $\alpha^{i}_{j} = \textbf{e}_c [G^{\#}_{v} \cdot i + j]$. Taking the popularity of items as input encourages the controller to assign larger importance weights to frequently visited item groups by local users, which are distinctive across different user groups with diversified user interests, further promoting personalization. The embedding block selection is based on the learned importance weights. In this manner, the task of proper embedding block search is altered to the importance weights $\boldsymbol{\alpha}$ parameters optimization by the controller.


\subsubsection{The optimization method}

\begin{algorithm}[!ht]
  \renewcommand{\algorithmicrequire}{\textbf{Input:}}
  \renewcommand{\algorithmicensure}{\textbf{Output:}}
  \caption{The PEEL Optimization Algorithm} \label{alg:PEEL}
  \begin{algorithmic}[1]
    \Require the user-item interaction instances and ground-truth labels
    \Ensure Refined RecSys parameters $\textbf{W}^{*}$ and well-learned controller parameters $\textbf{V}^{*}$
    \While{not converged} 
        \State Collect a mini-batch of validation data instances
        \State Estimate the approximation of $\textbf{W}^{*}(\textbf{V})$ via Eq.\ref{equa:gra}
        \State Update $\textbf{V}$ by descending  \par$\nabla_{\textbf{V}} \mathcal{L}_{val}(\textbf{W}-\xi\nabla_{\textbf{W}}\mathcal{L}_{train}(\textbf{W}, \textbf{V}), \textbf{V})$
        \State Sample a mini-batch of training data instances
        \State Generate $\boldsymbol{\alpha}$ via the controller with current $\textbf{V}$
        \State Generate prediction via the RecSys with current $\textbf{W}$ and $\boldsymbol{\alpha}$
        \State  Evaluate the prediction
        \State Update $\textbf{W}$ by descending $\nabla_{\textbf{W}}\mathcal{L}_{train}(\textbf{W}, \textbf{V})$
        
    \EndWhile

    \end{algorithmic}
    \label{algorithm:1}
\end{algorithm}

We look into the optimization of PEEL in this subsection. The optimization task is to jointly optimize the parameters of the group-level fine-tuning RecSys $\textbf{W}$, and the parameters of controllers $\textbf{V}$. We cannot first optimize the RecSys, then optimize the controller, since they are heavily interdependent. The controller's output is the importance weight in RecSys, and the loss of RecSys is the controller's input. Moreover, it is not applicable to train $\textbf{W}$ and $\textbf{V}$ on the same training batch due to the potential overfitting problem mentioned above. Therefore, we propose the PEEL optimization algorithm to learn the $\textbf{W}$ and $\textbf{V}$ in a bi-level optimization manner, inspired by neural architecture search \cite{liu2018darts}. We update $\textbf{V}$ based on validation loss $\mathcal{L}_{val}$  and update $\textbf{W}$ based on training loss $\mathcal{L}_{train}$ through back-propagation, which can be considered as a bi-level optimization problem \cite{pedregosa2016hyperparameter,pham2018efficient}:

\begin{subequations}
\begin{align}
& \min_{\textbf{V}} \mathcal{L}_{val}(\textbf{W}^*(\textbf{V}),\textbf{V}) \\
\label{equa:bi1}
& \text{s.t.} \ \textbf{W}^*(\textbf{V}) = \arg \min_{\textbf{W}}\mathcal{L}_{train}(\textbf{W},\textbf{V}^{*})
\end{align}
\label{equa:bi2}
\end{subequations}
where  $\textbf{W}$ is the lower-level parameter, and $\textbf{V}$ is the upper-level parameter. Training internal optimization fully to convergence is time-consuming and prohibitively expensive. Therefore,  we approximate $\textbf{W}^{*}(\textbf{V})$ by varying $\textbf{W}$ with one training step only \cite{liu2018darts}, which can be called as one-step approximation:
\begin{subequations}
\begin{align}
    & \nabla_{\textbf{V}}\mathcal{L}_{val}(\textbf{W}^*(\textbf{V}),\textbf{V}) \\
    \approx &\nabla_{\textbf{V}}\mathcal{L}_{val}(\textbf{W} - \xi \nabla_{\textbf{W}}\mathcal{L}_{train}(\textbf{W},\textbf{V}), \textbf{V})
\end{align}
\label{equa:gra}
\end{subequations}
where $\textbf{W}$ is the contemporary weights obtained by the optimizations, and $\xi$ indicates the learning rate for one step in the internal variable optimization. We present our optimization algorithm in Algorithm \ref{algorithm:1}. This optimization procedure is conducted on each user group independently on the server side in a one-off manner, resulting in parallel implementation.


\subsubsection{Cusotmizing PEEs with block-wise weights}
After the block-wise weights are fully optimized by the algorithm mentioned above. We construct the PEEs based on the weighted embedding blocks and deploy PEEs to the devices with a memory budget $M$. For instance, the on-device memory budget $M= 10MB, N=8, |\mathcal{V}|=50,000,\ d=16,$ and $G^{\#}_v=20$. Since each parameter occupies \SI{4e-6}{MB} in a 32-Bit floating point operation system, one embedding block with \SI{3.2e5} parameters occupies 0.16{MB}, the maximum number of selected embedding blocks $C= \lfloor \frac{M -size(\textbf{u}_{i})}{0.16MB} \rfloor = 62$. The embedding blocks with the highest importance weight within each item group are retained to guarantee that each item group has at least one EB. Then the rest $(C-G_v^{\#})$ EBs are selected w.r.t. the weights $\boldsymbol{\alpha}$ in descending order. For a user $u_i \in \mathcal{U}$, the PEE of item $v_j$ is represented as $\textbf{v}^{i}_{j}=[\hat{e}^j_n]_{n \in \mathcal{S}^{i}_{j}}$, the concatenation of the selected embedding blocks, and $|\mathcal{S}^{i}_{j}| \leq N$. In practice, we can customize the model based on the maximum memory budget allowed for each device, hence the deployed on-device recommender only needs to handle situations where the memory budget shrinks. Once the memory budget decreases after deployment, an on-device model can instantly deactivate less important blocks, thus retaining its accuracy under the constantly changing device capacity.

\subsubsection{Discussions.} 
 We hereby compare the technical backbone of PEEL with two major types of existing methods.
 (1) Reinforcement learning-based search \cite{10.1145/3397271.3401436,10.1145/3336191.3371858,joglekar2020neural}: The controller trained with non-differentiable reinforcement learning is only optimized for compressing the full embedding table towards a single configuration, while PEEL's end-to-end learned importance weights facilitate efficient adaptation to numerous memory budgets. (2) One-shot search \cite{rockyChenKDD21,qu2022single,cheng2020differentiable}: Like PEEL, they decouple the learning and resizing procedures of embedding tables, and do not require learning from scratch when given a different memory budget. However, their deployed recommenders lack an efficient mechanism to handle the dynamic on-device resource without calling the search function again. In contrast, once deployed by PEEL, the embeddings can self-adjust to suit stricter memory budgets in a fully on-device manner. Furthermore, all existing embedding customization methods are incapable of generating personalized on-device embedding tables, which is a key performance bottleneck tackled by PEEL.
 

\subsection{On-Device Ranking}
\label{sec:ranking}
After PEEs custmizing, the PEEs for all items $\{\textbf{v}^{i}_{j}\}_{ 1 \leq i \leq |\mathcal{V}|}$ and the user embedding $\textbf{u}_i$ are deployed on the device of user $u_i$ for on-device ranking. As for the ranking of conventional memory-efficient RecSys, the dot product \cite{johnson2019billion} or MLP with the concatenation of two embeddings \cite{10.1145/3038912.3052569} as the input can be utilized as the similarity function to estimate the similarity. However, the PEE of items and the user embeddings have mismatched dimensions, leading to infeasibility with the similarity functions mentioned above. Inspired by RULE \cite{rockyChenKDD21}, we utilize the similarity function shown below for on-device ranking:

\begin{equation}
r_{ij} = \frac{max(\{ \mathcal{S}^{i}_k\}_{1 \leq k \leq G_{v}^{\#}})}{|\mathcal{S}^{i}_{j}|} \sum_{n \in \mathcal{S}^{i}_{j}} \textbf{u}_{i}^{\top} \cdot \bar{\textbf{e}}^j_n
\end{equation}
where $\bar{\textbf{e}}^j_n \in \mathbb{R}^{D}$ is the repetition of ${\textbf{e}}^j_n$ for $N$ times, and the fraction part normalizes  embeddings from different item groups, and makes the similarity scores $r_{ij} $ magnitude-comparable among all items. This similarity function for on-device ranking does not require excessive parameters nor occupy memory budgets from PEEs. The space complexity for each device in the proposed method PEEL is $O(|\mathcal{V}| \cdot D + D)$, where $|\mathcal{V}|$ represents the number of items, and $D$ denotes the embedding size, as investigated in Section \ref{sec:exp_hs}. Importantly, for each device, it only stores its corresponding user embedding, and this space requirement is not influenced by the total number of users, denoted as $|\mathcal{U}|$.


\section{Experiments}
\label{sec:experiment}
To validate the effectiveness of PEEL\footnote{The code is released on https://anonymous.4open.science/r/PEEL/}, we conduct extensive experiments aiming to answer the following research questions (RQs):
\begin{itemize}
    \item \textbf{RQ1}: How does PEEL perform compared with other recommendation methods?
    \item \textbf{RQ2:} Can PEEL handle the dynamic memory budgets scenarios (i.e., the budgets for users vary over time)?
    \item \textbf{RQ3:} How do different hyper-parameters stated in Section \ref{sec:proposed method} influence the performance of PEEL?
    \item \textbf{RQ4:} How do different components of PEEL affect the recommendation performance?
    \item \textbf{RQ5:} What is the on-device usability of PEEL, particularly in terms of latency during deployment on heterogeneous devices?

\end{itemize}

\subsection{Experimental Settings}

\subsubsection{Datasets}
The experiments are conducted on two widely used public datasets, Amazon-Book \cite{Ups_and_Downs} and Yelp2020 \cite{rockyChenKDD21}. The Amazon-Book consists of Amazon users' preference for books and is the largest dataset in the crawled collection by \cite{Ups_and_Downs}. The Yelp2020, released by the open review platform Yelp, consists of users' reviews on various businesses. Both datasets, containing millions users/items interactions, are available to the public. Following \cite{rockyChenKDD21,10.1145/3038912.3052569}, we filter out those users and items with less than ten interactions, and both datasets are split into training, validation, and test with the ratio 7:1:2. The statistics are summarized in Table \ref{tab:dataset}.
\vspace{-5pt}
\begin{table}[htbp]
 \renewcommand{\arraystretch}{1}
  \centering
  
  \caption{The statistics of datasets.}
\begin{tabular}{c|c|c}
\toprule
Dataset & Amazon-Book & Yelp2020 \\
\midrule
\midrule
\#User & 52,643 & 138,322 \\
\#Item & 91,599 & 105,843 \\
\#Iteractions & 2,984,108 & 3,865,586 \\
\bottomrule
\end{tabular}%
  \label{tab:dataset}%
  \vspace{-10pt}
\end{table}%

\subsubsection{Baselines}
We compare PEEL with the following state-of-the-art methods, including fix-sized embedding methods, compositional embedding methods, multi-dimensional embedding methods, as well as the elastic embedding method.

\textbf{Fix-sized embedding:} \textbf{PMF \cite{mnih2007probabilistic}} learns the latent factors of items and users by the probability density function of the Gaussian distribution. \textbf{LightGCN \cite{he2020lightgcn}} develops a light graph convolution to learn the user and item embeddings via the neighborhood aggregation mechanism.

\textbf{Multi-dimensional embedding:} \textbf{ESAPN \cite{10.1145/3397271.3401436}} utilizes a reinforcement learning agent to automatically search the user/item embedding dimensions dynamically dependent on their popularity. \textbf{AutoEmb \cite{9679068}} proposes an end-to-end framework to select embedding dimensions for users and items.

\textbf{Compositional embedding:} \textbf{DLRM-CE \cite{10.1145/3394486.3403059}} outputs compositional embeddings with the quotient-remainder trick, reducing the embedding size while ensuring the uniqueness of each interaction category's representation. \textbf{LightRec \cite{10.1145/3366423.3380151}} utilizes parallel codebooks to compose the item embeddings.

\textbf{Elastic embedding with embedding blocks:} \textbf{RULE \cite{rockyChenKDD21}} proposes an on-device recommendation paradigm for various memory budgets for devices by learning full embeddings and searching item embeddings with an evolutionary algorithm.

\begin{table*}[htbp]
\renewcommand{\arraystretch}{0.85}
  \centering
  \caption{Recommendation results. Numbers in \textbf{bold} are the best results for corresponding metrics.}
  \begin{adjustbox}{max width=\textwidth}
\begin{tabular}{c|c|c|cccccc}
\toprule
\multirow{2}[4]{*}{Dataset} & \multicolumn{2}{c|}{\multirow{2}[4]{*}{Method}} & \multicolumn{2}{c|}{5MB} & \multicolumn{2}{c|}{10MB} & \multicolumn{2}{c}{25MB} \\
\cmidrule{4-9}      & \multicolumn{2}{c|}{} & \multicolumn{1}{c|}{Recall@50} & \multicolumn{1}{c|}{NDCG@50} & \multicolumn{1}{c|}{Recall@50} & \multicolumn{1}{c|}{NDCG@50} & \multicolumn{1}{c|}{Recall@50} & NDCG@50 \\
\midrule
\multirow{8}[8]{*}{Amazon-Book} & \multirow{2}[2]{*}{Fix-sized } & PMF   & 0.02552 & 0.02603 & 0.02590 & 0.02869 & 0.02894 & 0.03118 \\
      &       & LightGCN & 0.05257 & 0.09102 & 0.06284 & 0.10750 & 0.06475 & 0.13450 \\
\cmidrule{2-9}      & \multirow{2}[2]{*}{Compositional } & DLRM-CE & 0.02371 & 0.04033 & 0.02818 & 0.06018 & 0.03300 & 0.06948 \\
      &       & LightRec & 0.04408 & 0.07662 & 0.05045 & 0.07969 & 0.05921 & 0.10345 \\
\cmidrule{2-9}      & \multirow{2}[2]{*}{Multi-dimensional } & ESAPN & 0.02199 & 0.04040 & 0.02323 & 0.05125 & 0.02490 & 0.05891 \\
      &       & AutoEmb & 0.02035 & 0.04203 & 0.02284 & 0.05156 & 0.02403 & 0.05215 \\
\cmidrule{2-9}      & \multirow{2}[2]{*}{Elastic} & RULE  & 0.05334 & \textbf{0.09219} & 0.06512 & 0.11402 & 0.06602 & 0.12867 \\
      &       & \textbf{PEEL} & \textbf{0.05351} & 0.09122 & \textbf{0.06643} & \textbf{0.12167} & \textbf{0.06639} & \textbf{0.13962} \\
\midrule
\midrule
\multirow{8}[8]{*}{Yelp2020} & \multirow{2}[2]{*}{Fix-sized } & PMF   & 0.00597 & 0.01095 & 0.00849 & 0.01325 & 0.00950 & 0.01575 \\
      &       & LightGCN & 0.01549 & 0.01867 & 0.01804 & 0.02809 & 0.01893 & 0.03315 \\
\cmidrule{2-9}      & \multirow{2}[2]{*}{Compositional } & DLRM-CE & 0.01268 & 0.01230 & 0.01435 & 0.01627 & 0.01445 & 0.01953 \\
      &       & LightRec & 0.00768 & 0.01573 & 0.01255 & 0.01456 & 0.01350 & 0.02094 \\
\cmidrule{2-9}      & \multirow{2}[2]{*}{ Multi-dimensional } & ESAPN & 0.00698 & 0.01431 & 0.00838 & 0.01490 & 0.00874 & 0.01589 \\
      &       & AutoEmb & 0.00729 & 0.01353 & 0.00812 & 0.01437 & 0.00914 & 0.01471 \\
\cmidrule{2-9}      & \multirow{2}[2]{*}{Elastic} & RULE  & \textbf{0.01752} & 0.01945 & 0.02337 & 0.03236 & 0.02320 & 0.02881 \\
      &       & \textbf{PEEL} & 0.01571 & \textbf{0.02237} & \textbf{0.02466} & \textbf{0.03299} & \textbf{0.03455} & \textbf{0.03582} \\
\bottomrule
\end{tabular}%
    \end{adjustbox}
    \label{tab：results}
    \vspace{-5pt}
\end{table*}

\subsubsection{Evaluation Protocols}
Following \cite{chen2022fast,rockyChenKDD21,10.1145/3366423.3380187}, we use two evaluation metrics \textit{Recall@50} and \textit{NDCG@50} (Normalized Discounted Cumulative Gain) to measure model performance \cite{10.1145/3447548.3467097,he2020lightgcn,10.1145/3038912.3052569}. \textit{Recall@50} evaluates the proportion of the ground truth items included on the top-k list, and \textit{NDCG@50} measures whether the ground truth items can be ranked as highly as possible. Following \cite{rockyChenKDD21,he2020lightgcn}, we average results over all users, and all items that have no interaction with the users are used as negative samples for recommendation evaluation.

\subsubsection{Hyper-parameters Setting}   
The hyper-parameters of the proposed method PEEL are set as follows. The number of groups for users $G_u^{\#}$ and the number of groups for items $G_v^{\#}$ are set as 15 and 20. the length of embedding blocks $d$ and number of embedding blocks for one item $N$ is set as 8 and 16, respectively. By default, PEEL segments items into groups with equal size according to the item popularity, and clusters users by the $K$-Means \cite{grira2004unsupervised}. The effectiveness of item popularity segmentation and user clustering will be investigated in Section \ref{sec:ablation}. All the baselines are implemented by the codes provided by the authors. Adam optimizer \cite{kingma2014adam} with a learning rate of 0.001 and early stopping \cite{zhu2021open} are utilized.

\subsection{Top-k Recommendation (RQ1)}

To validate the effectiveness of PEEL, we compare it with baselines on the Top-k recommendation task. Following \cite{rockyChenKDD21}, we test all methods under  budget  $M$, where $M$
 is set as 5MB, 10MB, and 25MB. For every method, the size of the embedding tables, weights, and biases are adjusted to fit the memory budget. From the experimental results shown in Table \ref{tab：results}, we can observe that:

     All models perform better on both datasets as the memory budget increases. It suggests that larger embedding tables show information expressiveness, resulting in better recommendation performance. Since the size of embedding tables plays an essential role in recommendation results, it is necessary to compare the performance of recommender systems under different memory budgets.

    RULE and PEEL perform better than multi-dimensional methods. The possible explanation is that multi-dimensional methods develop only one cloud-based model with automated designed user/ item embeddings without considering the limited memory budget on the device side. They pay attention to assigning different embedding dimensions to different feature fields, rather than effectively utilizing the limited embedding budgets, leading to unsatisfied results under on-device scenarios. The necessity and effectiveness of on-device budget-aware methods have been underlined.

    PEEL achieves better performance than the compositional embedding methods and fix-sized embedding methods. It indicates the effectiveness of elastic embeddings, which learn diverse embedding blocks for items. Although LightGCN has fixed-sized embeddings, it shows competitive results under different device budgets, underlining its generalization ability.

    PEEL outperforms all baselines in most cases on both datasets. Although RULE partially achieves better performance concerning NDCG@50 and Recall@50 under the 5MB device budget, the difference between PEEL and RULE is marginal, and PEEL performs the second best results among all methods. PEEL only searches under the maximum budget (i.e., 25MB), and generates satisfying results, while all baselines including RULE need retraining or reconfiguring when facing a new budget.  PEEL only requires one search under the maximum possible budget and can seamlessly change embedding tables according to varying memory constraints. This strength will be further investigated in Section \ref{sec:dynamic}.

\begin{figure*}
\begin{minipage}[t]{1.3\columnwidth}
  \includegraphics[width=\linewidth]{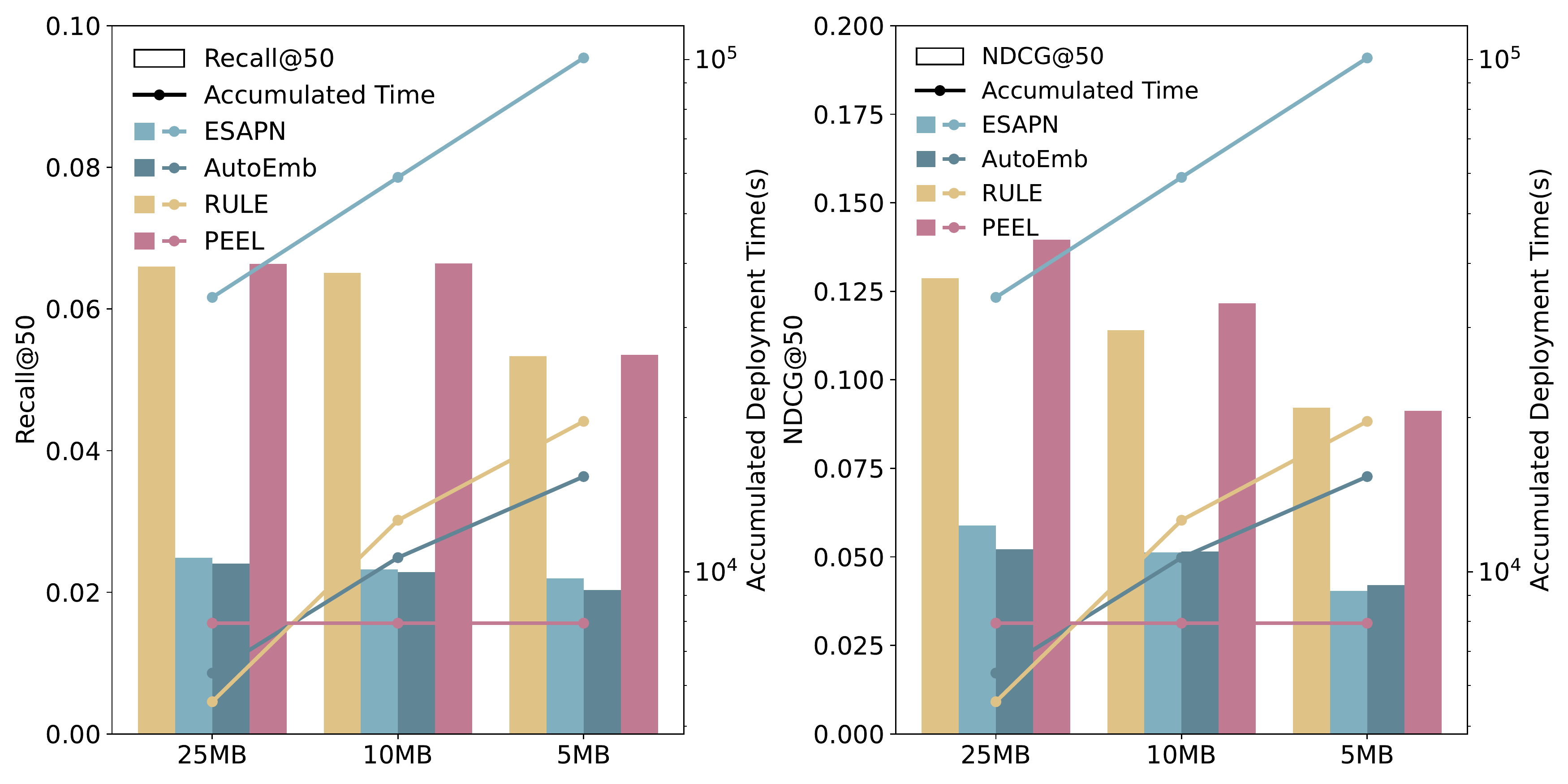}
  \caption{The performance and deployment time for PEEL and baselines on Amazon-Books under dynamic budget scenarios.}
    \label{fig:dynamic}
\end{minipage}\hfill
\begin{minipage}[t]{0.63\columnwidth}
  \includegraphics[width=\linewidth]{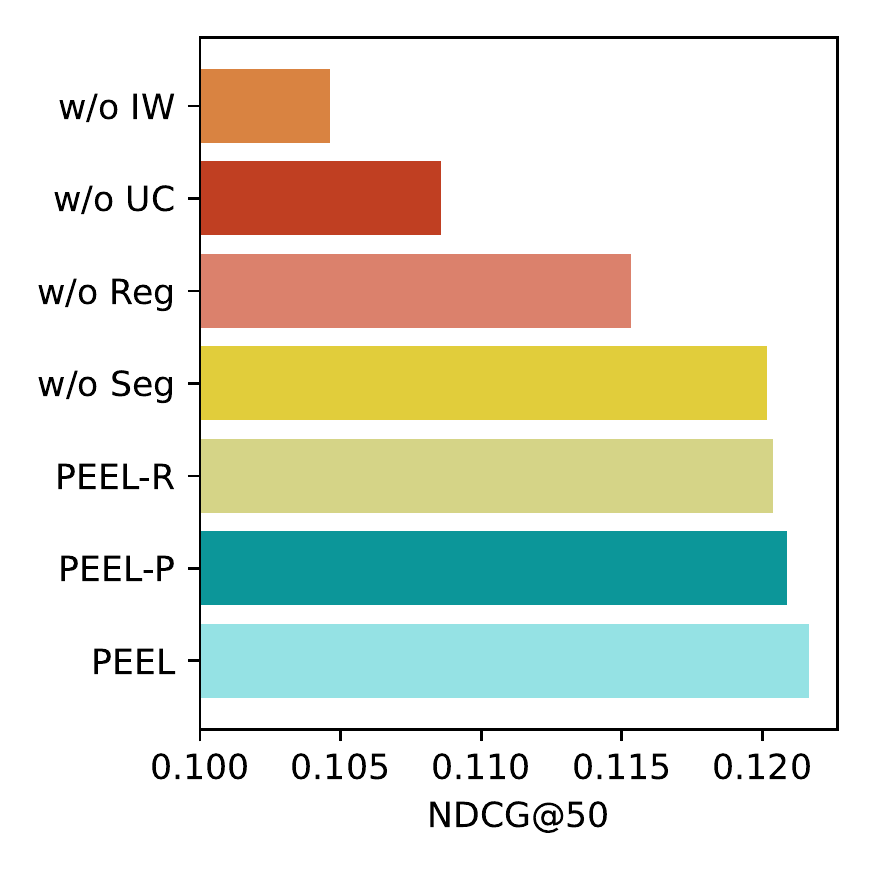}
  \caption{Result of ablation experiment on different variants of PEEL.}
\end{minipage}
\vspace{-5pt}
\end{figure*}

\subsection{Dynamic Device Budgets (RQ2)}
\label{sec:dynamic}

The on-device recommendations are the primary application of our proposed recommendation model. Although multi-dimensional methods and RULE can be implemented in on-device scenarios, they neglect the dynamic on-device budgets.

In practice, we customize the model based on the maximum memory budget allowed for each device, hence the deployed on-device recommender only needs to handle situations where the memory budget shrinks. To simulate the dynamic device budgets scenario, we first set the budget to 25MB, and record the recommendation performance, as well as the accumulated deployment time required to obtain the embedding tables. Then, the budget is set to 10MB and 5MB. The accumulated deployment time and performance for each memory budget are recorded in Figure \ref{fig:dynamic}. To ensure an equitable and thorough assessment of the model's performance when subjected to various memory budgets, each memory budget is analyzed under the same evaluation criteria over all items.

For all baselines, the accumulated deployment time grows with decreasing of dynamic memory budgets, while the deployment time for PEEL will not increase after the first deployment. Thousand seconds are essential for all baselines facing a new memory budget, which is impractical for real-world applications with instant response requirements. When the memory budgets shrink, the original embedding tables for baselines are not feasible and require training from scratch or searching for the proper embedding tables again. As for PEEL, once the weighted embedding blocks are learned over the full embedding tables, it can instantly discard unessential blocks on the device side as the budget decreases, without additional learning time or the help of a cloud server. Thus, PEEL is effective and efficient for dynamic memory budgets.

\vspace{-4pt}
\subsection{Hyperparameter Study (RQ3)}
\label{sec:exp_hs}
\begin{figure*}
\begin{minipage}[t]{1.2\columnwidth}
  \includegraphics[width=\linewidth]{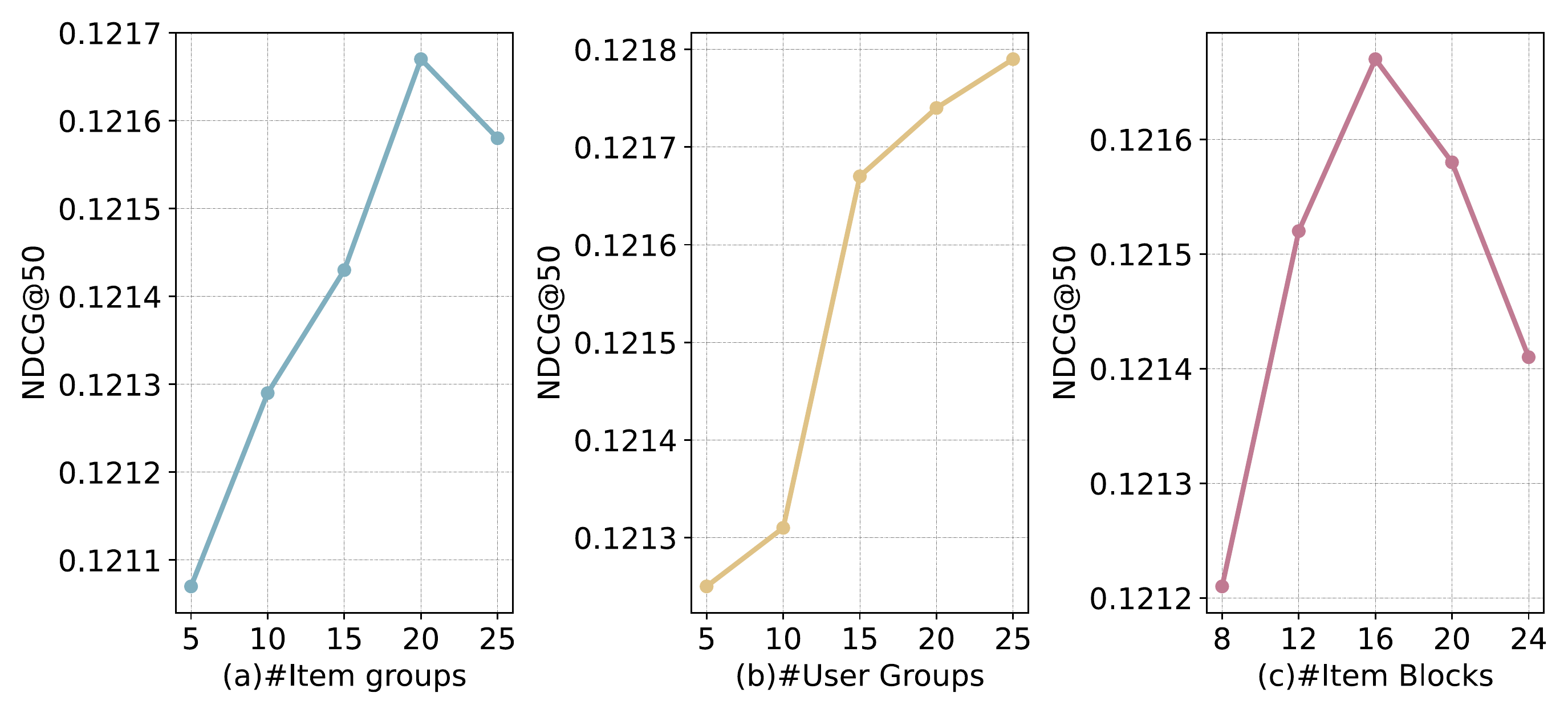}
  \caption{The performance of PEEL with different (a) The number of item segmentation groups (b) The number of user clustering groups. (c) The number of embedding blocks for each item.}
  \label{fig:hyper}
\end{minipage}\hfill 
\begin{minipage}[t]{0.78\columnwidth}
  \includegraphics[width=\linewidth]{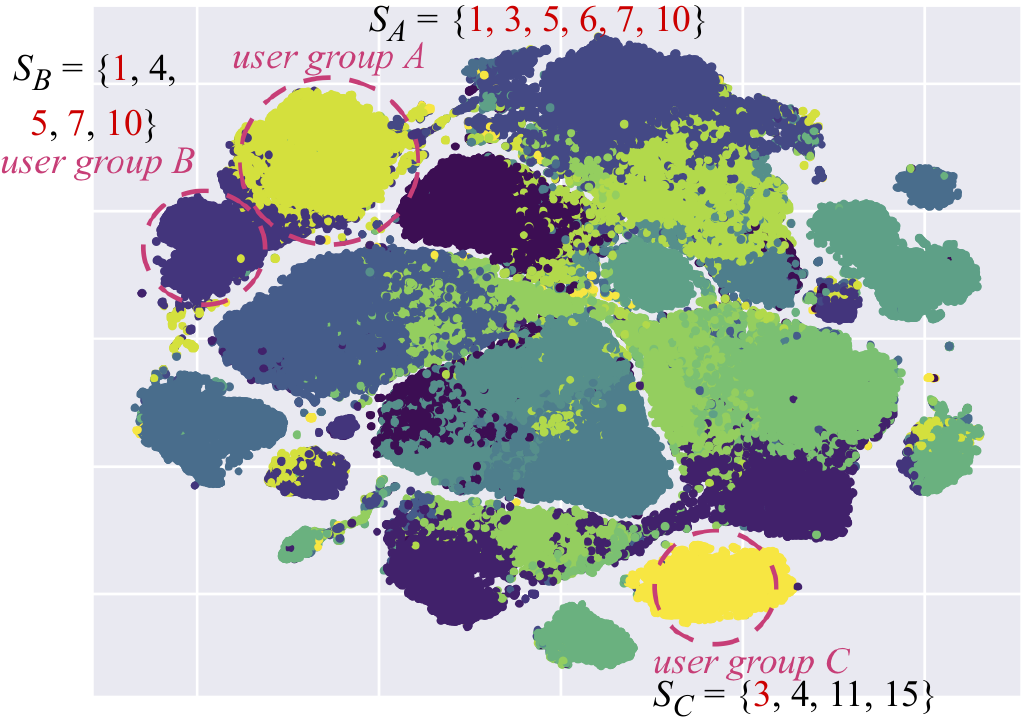}
  \caption{Visualization of user clustering, and indexes of selected EBs for one item segmentation. The user group number is 15.}
  \label{fig:visual}
\end{minipage}
\vspace{-15pt}
\end{figure*}

To explore the effect of three crucial hyper-parameters, namely
the number of item segmentation groups $G_v^{\#} = \{ 5, 10, 15, 20, 25\}$, the number of user clustering groups $G_{u}^{\#} = \{5, 10, 15, 20, 25 \}$, and the number of embedding blocks $N = \{8, 12, 16, 20, 24\}$, we conduct experiments on Amazon-Book under 10MB budget, and report performance with different hyper-parameters in Figure \ref{fig:hyper}.

\textbf{Impact of $G_v^{\#}$}. $G_v^{\#}$ determines the granularity of item segmentation. When the item groups are too small, distinctive items share the same embedding blocks and similar PEEs, resulting in unsatisfying performance. With the increase of $G_v^{\#}$, a finer granularity leads to better performance. When $G_v^{\#}$ reaches 25, the decreased performance may be caused by the fact that excessive
 embedding blocks lead to more difficulties in finding important blocks.

\textbf{Impact of $G_u^{\#}$}. $G_u^{\#}$ determines the number of group-level refined RecSys and controllers. Even though they can be conducted in parallel, increasing user groups introduce high computation costs on the cloud. More user groups generally contribute to higher recommendation accuracy owing to the higher personalization. When $G_u^{\#}$ reaches 15, the performance improvement tends to be marginal. One possible explanation is that 15 groups are expressive enough to include the diversified action tendencies among all users.

\textbf{Impact of $N$}. The length of embedding blocks $d = 8$ stays unchanged, leading to a full embedding dimension $D$ varies in $\{32, 64, 96, 128, 160\}$. Generally, more embedding block candidates $N$ lead to better performance. When the number of item blocks reaches 16, the performance decrease. The possible solution is that excessive item blocks bring more difficulties in finding proper PEEs.

\vspace{-4pt}
\subsection{Ablation Study (RQ4)}
\label{sec:ablation}
In this section, we aim to demonstrate the effect of item segmentation, diversity-driven regularizer, user clustering, and importance weight. In particular, we implement PEEL without one component, while keeping other components unchanged.  The experiments are carried out on the Amazon-Book dataset with a 10MB memory budget for illustration, and similar patterns emerge under different memory constraints and across other datasets

$PEEL-P$ utilizes the average pooling of all layers’ outputs instead of the final layer embedding. One possible explanation is the regularizer and the subsequent embedding selection.

$PEEL-R$ randomly segments items into groups instead of sorting them. The performance decreases emphasize the efficiency of popularity-based item segmentation. 

$w/o \  PopSeg$ discards the item popularity segmentation, and splits items randomly. The performance decreases because irrelevant items share the same embedding blocks and similar PEEs. 

$w/o \ Regularizer$ deletes the regularization term in Equation \ref{equa:loss_funcion}. One possible explanation for the performance drop is that the embedding blocks are not diversified nor expressive enough to capture the item representation. 

$w/o \ User \ Clustering$ discards the group-level fine-tuning strategy and generates the same model for all users. The performance considerably decreases, which validates the necessity of personalization. Moreover, Figure \ref{fig:visual} shows the visualization of 15 clustering groups of users under the 10MB budget for the Amazon-Book dataset. We empirically find that the number of the same searched blocks between two Groups increases when two groups are close (i.e., Group A and Group B) rather than far away (i.e., Group A and Group C). This finding can further emphasize the necessity of personalization and the effectiveness of clustering. 

$w/o \ Importance \ Weight$ deletes the controllers, and randomly selects embedding blocks, leading to a dramatic performance drop. Without importance weight indicating the contribution of each embedding block, random selection cannot generate effective PEEs, which indicates the effectiveness of importance weights.

\begin{table}[htbp]

\renewcommand{\arraystretch}{0.85}
  \centering
\caption{Average inference time of PEEL for each user.}
\begin{adjustbox}{max width=0.47\textwidth}
\begin{tabular}{c|cccccc}
\toprule
\multirow{2}[4]{*}{Dataset} & \multicolumn{3}{c|}{IoT Latency (ms)} & \multicolumn{3}{c}{Cellphone Latency (ms)} \\
\cmidrule{2-7}      & \multicolumn{1}{c|}{5MB} & \multicolumn{1}{c|}{10MB} & \multicolumn{1}{c|}{25MB} & \multicolumn{1}{c|}{5MB} & \multicolumn{1}{c|}{10MB} & 25MB \\
\midrule
\midrule
Amazon-Book & 352.17 & 576.15 & 886.72 & 71.89 & 154.97 & 214.39 \\
Yelp2020 & 390.64 & 754.99 & 943.43 & 85.26 & 185.28 & 250.78 \\
\bottomrule
\end{tabular}%
\end{adjustbox}
\label{tbl:inference}
\vspace{-10pt}
\end{table}

\subsection{Latency on Heterogeneous Devices (RQ5)}
Given the primary application of our proposed recommendation paradigm centered on device-side utilization, the real-life practicality of PEEL is evaluated by specifically assessing the latency on two distinct hardware configurations. By deploying PEEL within simulated IoT and cellphone environments using a Linux virtual machine, we determine the average inference time required to generate a Top-50 ranked item list for individual users. The on-device settings and latency of PEEL are subsequently described below.

\textbf{IoT: 1× vCPU (Intel i5-5350U) 64MB RAM. } It should be noted that, when considering configurations for both Internet of Things (IoT) devices and cellphones, practical RAM size is selected instead of the maximum memory allocation (specifically, 25MB) in order to accommodate the multitasking requirements commonly encountered in real-world scenarios. As shown in Table \ref{tbl:inference}, operating with a compact embedding size of under 5MB, the preparation of the item ranking list is accomplished in less than 200MS. Even as the embedding size expands to 25MB on the Yelp2020 dataset, PEEL consistently achieves an inference time below the 1000MS threshold.

\textbf{Cellphone: 4× vCPU (Intel i5-5350U) 512MB RAM.} { Equipped with an additional three CPUs compared to IoT devices, PEEL showcases negligible latency while conducting recommendations in a cellphone environment. Leveraging a compact embedding size of under 5MB, the preparation of the item ranking list requires less than 100MS. As the memory budget escalates from 5MB to 25MB, the execution time rises to 214.39MS and 250.78MS on the Amazon-Book and Yelp2020 datasets, respectively. Our approach presents an efficient and effective solution for diverse on-device recommendation tasks.}

\vspace{-8pt}
\section{Conclusion}
\label{sec:conclusion}
{
In this paper, we propose a new paradigm for on-device recommendation, which identifies the optimal PEEs for on-device recommendation under varying memory
budgets in one shot, without retraining or reconfiguring. Embedding blocks with importance weights enables the PEEL to be feasible for any arbitrary on-device memory budgets. Regarding the dynamic device budget, an on-device model can instantly deactivate less important blocks, thus retaining its accuracy under the
constantly changing device capacity.
A group-level fine-tuning strategy with personalized elastic embeddings addressed user heterogeneity by deploying a group-based personalized model for distinctive users. Extensive experiments verify that PEEL offers strong guarantees on the recommendation performance and efficiency under heterogeneous and dynamic budgets.}

\section*{Acknowledgement}
This work was supported by the Australian Research
Council Future Fellowship (Grant No. FT210100624), the
Discovery Project (Grant No. DP190101985), the National Natural Science Foundation of China (Grant No.
61761136008), the Shenzhen Fundamental Research Program (Grant No. JCYJ20200109141235597), the Guangdong
Basic and Applied Basic Research Foundation (Grant No.
2021A1515110024), the Shenzhen Peacock Plan (Grant No.
KQTD2016112514355531), the Program for Guangdong Introducing Innovative and Entrepreneurial Teams (Grant No.
2017ZT07X386).




%

\bibliographystyle{IEEEtran}

\bibliography{main}
%



%

\begin{IEEEbiography}[{\includegraphics[width=1in,height=1.25in,clip,keepaspectratio]{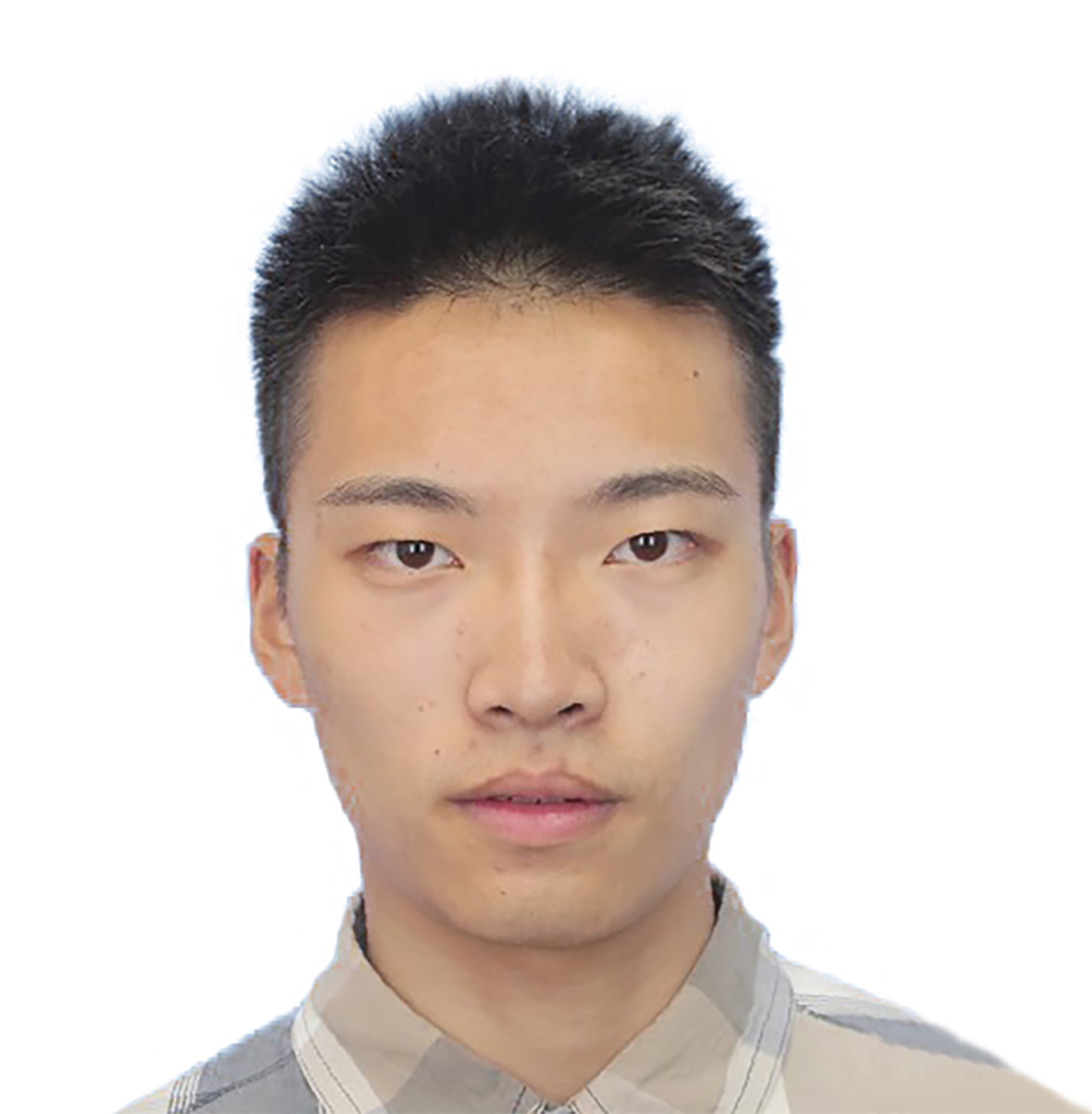}}]{Ruiqi Zheng} received a B.E. in Computer Science and Technology in 2021 from Southern University of Science and Technology. He is currently pursuing the Ph.D. degree under the jointly program between the University of Queensland, Brisbane, Australia, and the Southern University of Science and Technology, Shenzhen, China. His research interests include automated machine learning, and recommender system.

\end{IEEEbiography}

\begin{IEEEbiography}[{\includegraphics[width=1in,height=1.25in,clip,keepaspectratio]{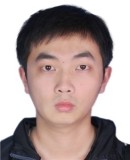}}]{Liang Qu} received a B.E. in Applied Physics in 2017, and an M.S. in Computer Science in 2019, from South China University of Technology and Harbin Institute of Technology respectively. He is currently pursuing the Ph.D. degree under the jointly program between the University of Queensland, Brisbane, Australia, and the Southern University of Science and Technology, Shenzhen, China. His research interests include swarm learning, recommender system, and graph embedding.

\end{IEEEbiography}

\begin{IEEEbiography}[{\includegraphics[width=1in,height=1.25in,clip,keepaspectratio]{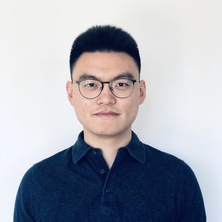}}]{Tong Chen} received the Ph.D. degree in computer science from the University of Queensland in 2020. He is a lecturer and ARC DECRA Fellow with the School of ITEE, The University of Queensland. His research work has been published on top venues like SIGIR, SIGKDD, ICDE, WWW, ICDM, IJCAI, AAAI, CIKM, TOIS, TKDE, etc., where his research interests include data mining, machine learning, recommender systems, and predictive analytics.
\end{IEEEbiography}

\begin{IEEEbiography}[{\includegraphics[width=1in,height=1.25in,clip,keepaspectratio]{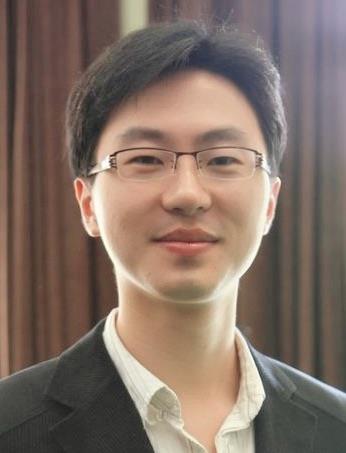}}]{Kai Zheng} is a Professor of Computer Science with University of Electronic Science and Technology of China. He received his PhD degree in Computer Science from The University of Queensland in 2012. He has been working in the area of spatial-temporal databases, uncertain databases, social-media analysis, inmemory computing and blockchain technologies. He has published over 100 papers in prestigious journals and conferences in data management field such as SIGMOD, ICDE, VLDB Journal, ACM Transactions and IEEE Transactions. He is a senior member of IEEE.
\end{IEEEbiography}


\begin{IEEEbiography}[{\includegraphics[width=1in,height=1.25in,clip,keepaspectratio]{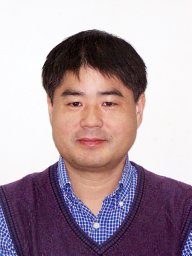}}]{Yuhui Shi} (Fellow, IEEE) received the Ph.D. degree in electronic engineering from Southeast
University, Nanjing, China, in 1992. He is currently a Chair Professor with the Department of Computer Science and Engineering, Southern University of Science and Technology, Shenzhen, China. He has coauthored the book \emph{Swarm Intelligence} (with Dr. J. Kennedy and Prof. R. Eberhart) and another book \emph{Computational Intelligence: Concept to Implementation} (with Prof. R. Eberhart). His main research interests are artificial intelligence, particularly swarm intelligence.
\end{IEEEbiography}

\begin{IEEEbiography}[{\includegraphics[width=1in,height=1.25in,clip,keepaspectratio]{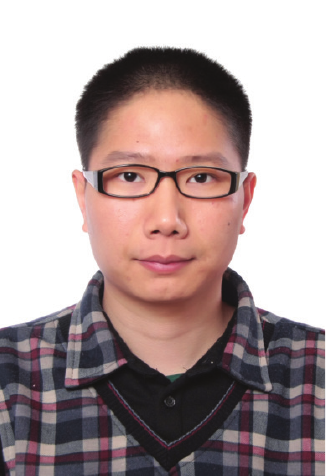}}]{Hongzhi Yin} received a PhD degree in computer science from Peking University, in 2014. He works as an ARC Future Fellow, Full Professor, and Director of the Responsible Big Data Intelligence Lab (RBDI) at The University of Queensland, Australia. He has made notable contributions to predictive analytics, recommendation systems,  graph learning, and decentralized and edge intelligence. He has published 260+ papers with an H-index of 66 and received eight best paper awards/runner-ups at the top conferences such as ICDE 2019, WSDM 2023, and DASFAA 2020. 
\end{IEEEbiography}

\end{document}